\documentstyle[twocolumn,prb,aps]{revtex}
\begin{document}
\draft
\title{Magnetic Phase Diagram
       of the Ferromagnetically Stacked Triangular XY Antiferromagnet:
       A Finite-Size Scaling Study }
\author{M.L. Plumer, A. Mailhot,\cite{note} and A. Caill\'e}
\address{ Centre de Recherche en Physique du Solide et D\'epartement de
Physique}
\address{Universit\'e de Sherbrooke, Sherbrooke, Qu\'ebec, Canada J1K 2R1}
\date{December 1993}
\maketitle
\begin{abstract}
Histogram Monte-Carlo simulation results are presented for the
magnetic-field -- temperature phase diagram of the XY model on a stacked
triangular lattice with antiferromagnetic intraplane and ferromagnetic
interplane interactions.   Finite-size scaling results at
the various transition boundaries are consistent with
expectations based on symmetry arguments.  Although a molecular-field
treatment of the Hamiltonian fails to reproduce the correct structure
for the phase diagram, it is demonstrated that a phenomenological Landau-type
free-energy model contains all the esstential features.  These results serve to
complement and extend our earlier work [Phys. Rev. B {\bf 48}, 3840 (1993)].
\end{abstract}
\section{Introduction}

Although finite-size scaling studies of critical phenomena based on
Monte-Carlo histogram (MCH)
data\cite{ferr} for unfrustrated systems has proven highly
effective in the estimation of critical exponents\cite{expon}
and in the determination of weakly first-order transitions,\cite{first}
corresponding studies of possibly more interesting frustrated lattices
have proven more demanding.\cite{frust,ising,mail,plum,rev}  The utility of
this
approach in examining the critical properties of the variety of
phase transitions which occur under the influence of an applied magnetic field
in the ferromagnetically stacked triangular XY antiferromagnet is demonstrated
here.  The Hamiltonian is written as
\begin{equation}
{\cal H}~=~J_{\|} \sum_{<ij>} {\bf S}_i \cdot {\bf S}_j
+ J_{\bot} \sum_{<kl>} {\bf S}_k \cdot {\bf S}_l
- H \sum_i S_{xi}
\end{equation}
where the spins lie in the basal plane, $J_\|<0$ is the ferromagnetic
interplane interaction, $J_{\bot}>0$ indicates
the antiferromagnetic coupling which is frustrated for the triangular
geometry, $<i,j>$
and $<k,l>$ represent near-neighbor sums along the hexagonal $c$ axis
and in the basal plane, respectively, and the field is applied
in the basal-plane direction $x$.
The general structure of the phase diagram shown in Fig. 1 was
determined in our earlier work\cite{plum} (hereafter referred to as I).
Phases are labelled by the nonzero components of the (complex) spin
polarization vector ${\bf S} = {\bf S}_a + i{\bf S}_b$.  The N\'eel transition
at zero field is to the well-known helically polarized $120^\circ$ spin
structure ($S_{ax}=S_{by}$).  Indicated on the figure are the elliptically
polarized phase $7$, linear phases $6$ and $9$, as well as the paramagnetic
state $1$.  Phase $6$ is of particular interest
as it has the symmetry of the 3-state Potts model and should exhibit a weak
first-order transition to the paramagnetic state.  This was confirmed by
extensive finite-size scaling of the extrema in various thermodynamic
functions which occur at the $1-6$ phase boundary for magnetic field
strengths $H=0.7$ and $H=1.5$.  Data for $H=0.7$ can be found in I.  In
addition to presenting the corresponding MCH data at $H=1.5$, finite-size
scaling results are given here for the other three transition lines, at the
points indicated on Fig. 1.

It is of interest to note that Lee {\it et al.}\cite{lee}
examined the XY antiferromagnet on a triangular lattice (un-stacked) in an
applied magnetic field.  At H=0, the transition shows Kosterlitz-Thouless
(KT) behavior but the field breaks rotational symmetry and
transitions involving true long-range spin order can occur.   The resulting
phase diagram in this 2D case is very similar to that of Fig. 1.  Using
traditional finite-size scaling of their MC data, these authors reported
that the $1-6$ transition belongs to the 2D 3-state Potts universality class,
but that the other transition lines exhibited nonuniversal critical behavior;
different exponents were found for different points on a transition line.
We find this latter conclusion is somewhat surprising but may be due to KT-like
excitations in some cases.  The possibility always exists that the
MC data were not sufficiently accurate to distinguish among the variety of
possibilities.

Before discussing our own MCH data for the 3D model, it is useful to
examine the results of a mean-field analysis based on the Landau theory of
phase transitions.


\section{Mean-Field Theory}

Mean-field analyses of magnetic phase diagrams based on Landau-type free
energies for frustrated spin systems have proven to be quite successful
in reproducing the essential features of both MC and experimental
results.\cite{rev}  (A noteable exception is the XY model on a
stacked triangular lattice, with both $J_\|, J_{\bot} > 0$,
in the quasi-2D case\cite{plum2} where $J_\|<<J_{\bot}$).
It was previously demonstrated for the 2D version of the present model
that a molecular-field
treatment of the Hamiltonian (1) yields a phase diagram with phase $6$
absent.\cite{lee2}  Identical results are expected for the 3D model
under consideration here.  It is shown below, however, that the
phenomenological
Landau approach can capture all of the essential features of the phase
diagram of Fig. 1.  Such a treatment is also useful in understanding
analytically the interactions which are responsible for stabilizing each phase.
In addition, an examination of such a
free energy (which has the same structure as an appropriate
Landau-Ginzburg-Wilson Hamiltonian), together with symmetry arguments,
is useful in determining expectations
regarding the critical behavior of the various transition lines.

Following the method outlined in Ref.\onlinecite{plum3} , the free
energy is expanded to fourth order in the spin density
\begin{equation}
{\bf s(r)}~=~{\bf m} + {\bf S}e^{i{\bf Q \cdot r}}
                     + {\bf S^{\ast}}e^{-i{\bf Q \cdot r}}
\end{equation}
where {\bf m} is the uniform component induced by the magnetic field and
{\bf Q} is the wave vector.  The result can be written as
\begin{eqnarray}
F&~=~& A_Q S^2 ~+~ {\textstyle \frac12}A_0'm^2
\nonumber \\
&~+~& {\textstyle \frac14} B_3 m^4 ~+~ 2 B_4 \mid {\bf m} \cdot {\bf S} \mid^2
\nonumber \\
&~+~& B_6 [({\bf m} \cdot {\bf S})({\bf S} \cdot {\bf S})
\nonumber \\
&~+~& \cdot \cdot \cdot~-~ {\bf m} \cdot {\bf H},
\end{eqnarray}
where $A_Q = a(T - T_Q)$ and $A_0' = a(T - T_0')$.  This expression is
identical to that used in our analysis of the magnetic phase diagram
for the case of antiferromagnetic interplane coupling\cite{plum4}
{\it except} for the additional term $B_6$.  As emphasized in I,
this term {\it cubic} in ${\bf S}$ occurs since the ordering wave vector
satisfies the relation $3{\bf Q} = {\bf G}$, where ${\bf G}$ is a reciprocal
lattice vector.  The cubic term is responsible for the stability of the
3-state Potts phase $6$.  It is important to note that from symmetry
arguments alone, each of the fourth
order coefficients $B_i$ in (3) is independent since each corresponding
term is an independent invariant with respect to the relevant symmetry
operations.

A free energy identical in structure to (3)
may also be {\it derived} from a molecular-field
treatment of the Hamiltonian (1).\cite{plum5}  In this case, all the
fourth-order coefficients are the same and given by $B_i=bT$, where
for classical spins $b=\frac95$.  Since the Landau expansion is
applicable only in regions where $S$ is small, the approximation
$B_i \simeq bT_N$ is usually made.  In addition, the molecular-field
treatment yields $a=3$, as well as
\begin{equation}
T_Q~=~2(-J_\| + 3J_\bot)/a,~~~T_0'~=~-2(J_\| + 3J_\bot)/a.
\end{equation}
Thus, the exchange constants are the only parameters which appear in
this theory.

Fig. 2a shows the resulting phase diagram from an analysis of the
free energy from molecular-field theory expanded to sixth order in ${\bf s}$
using $J_\|=-1$, $J_\bot=1$ and ${\bf H} \| {\bf \hat x}$.  It is seen to
have the same structure as the mean-field result in Ref.\onlinecite{lee2}.
The 3-state Potts
phase $6$ does not appear.  An analysis of the free energy reveals that
this state is somewhat accidentally excluded.  This conclusion is
demonstrated by results from the more general phenomenological
model.  Analysis was made using the same parameter-values as in
the molecular-field theory, except that one of the fourth-order
coefficients $B_i$ was made to be different from $bT$.
Results for the cases $B_4 \equiv 1$ and 0.1 are shown in Figs.
2b and 2c, respectively.  Phase $6$ now appears.  The correct structure
of Fig. 1 is reproduced by the smaller value of $B_4$.  Similar results
occur if only $B_2$ or $B_3$ are set to be different from the other
$B_i$. It seems that
some effects of critical fluctuations not accounted for within
mean-field theory can be mimicked by the more general phenomenological model
in this case.

The stability of each state in the phase diagram can be understood by
examining the terms in (3) with the assumption $B_i > 0$.  At zero
applied field ($m=0$), the $B_2$-term is minimized with
${\bf S} \cdot {\bf S} = 0$.  This is achieved by a helical polarization
$S_{ax}=S_{by}$.  At low temperatures and low values of the applied field,
the $B_4$-term, which favors a configuration ${\bf S} \bot {\bf m}$,
distorts the helix into an ellipse (phase $7$), $S_{ax} \not= S_{by}$.
At higher temperatures and low field values,
the $B_6$-term dominates and favors phase $6$ with a configuration
${\bf S} \| {\bf m}$.  The high-field phase $9$ is a result of both
$B_4$ and $B_6$-type interactions which favor a linear polarization.

This analysis allows some predictions to be made regarding the nature
of the phase transition lines of Fig. 1.  The $1-6$ line should be
first order due to the relevance of the cubic term $B_6$, as confirmed in I.
(We note, however, that mean-field theory would also suggest the same behavior
in 2D.  In fact, this transition belongs to the 2D 3-state Potts universality
class.)
A continuous $1-9$ transition is the result of the mean-field analysis despite
the nonzero value of the Potts variable $S_{ax}$.  A first-order transition
does not occur in this case because, as discussed in I,
the $B_6$-term effectively disappears
at the paramagnetic boundary line at some power of $S$ greater than three.
It is not clear what the effects of critical fluctuations will be regarding
this scenario.  If the transition is continuous, it should belong to the
XY universality class since both $S_{ax}$ and $S_{ay}$ are involved.
The remaining two transition lines, $6-7$ and $6-9$ should both belong
to the Ising universality class as only a single component of ${\bf S}$
is involved in each case.  A purpose of the finite-size scaling analysis
described below is to test these predictions.

\section{Finite-Size Scaling}

The analysis of MC generated histograms used here
to determine finite-size scaling behavior is described in I.
In most cases, scaling was performed
on the extrema of a variety of thermodynamic functions, including the
specific heat $C$, susceptibility $\chi$,
energy cumulant $U(T)~=~1 - \frac13 \langle E^4 \rangle /
\langle E^2 \rangle ^2$, and
the logarithmic derivative of the order parameter $M$,
which is equivalent to
$V(T)~=~ \langle ME \rangle / \langle M \rangle  - \langle E \rangle $.
The energy cumulant exhibits a minimum near $T_N$, which achieves the value
$U^\ast = \frac23$ in the limit $L \rightarrow \infty$ for a continuous
transition, whereas $U^\ast < \frac23$
is expected in the case of a first-order transition.
The thermodynamic quantities should display volume-dependent scaling, $L^3$,
in the case of a first-order transition (except for $M$), or scaling as
$L^x$, where $x$ is a ratio of critical exponents, in the case of a
continuous transition.
Finite-size scaling results for the order parameter,
evaluated at the estimated critical temperature, are presented only
for the $1-9$ transition.
For reasons unknown to us, it appeared that the values of $L$ used here were
to small to obtain reliable results for this quantity in the other cases where
a continuous transition was expected.
Partly due to the relatively large fluctuations in the
MCH data, a general feature of frustrated systems,\cite{rev} the approach
taken in this work is to simply determine if the results are consistent
with the expectations and possibilities outlined above: Scaling was performed
with assumed exponent values.  This method of
presenting data was also used in I, as well as in Refs.\onlinecite{ising}
and \onlinecite{mail},
and is useful in cases where the true critical behavior may be revealed
only at larger lattice sizes.  The alternate approach of presenting data
in the form of Log-Log plots is more appropriate in cases where very reliable
statistics are available and finite-size correction terms to simple
$L^x$ behavior should be negligible.

Simulations were performed on the Hamiltonian (1) with
$J_\|=-1$ and $J_\bot=1$ using
periodic boundary conditions on $L \times L \times L$ lattices.
In most cases, a random initial spin configuraion was used and
thermodynamic averages were estimated after discarding
the first $2 \times 10^5$ MC steps for thermalization.  In other cases,
the final, well thermalized, configuration of a previous run was used
for the initial spin directions.  Further details are given in the following
sections which describe the results of finite-size scaling at
each phase-boundary line.

\subsection{H=1.5: $1-6$ Transition}

Results of the histogram analysis at $H=1.5$ for the $1-6$ phase boundary
are also discussed briefly in I.  Simulations were performed on lattices
$L=12-33$ with $1-2.6 \times 10^6$ MC steps used for thermal averaging.
The increased number of MC steps for larger lattices was used in an effort
to account for an expected increase in the correlation time.  Histograms
were made at $T=1.52$ and $T=1.523$ and the critical temperature was estimated
to be $T_c = 1.522(2)$ from the finite-size dependence of the various
thermodynamic extrema.  The results shown in Fig. 3 demonstrate that the
expected volume dependence of a first-order transition may not be evident
at smaller values of $L$.  This aspect of MC simulations has only recently been
emphasized.\cite{first,mail}  The extrapolated value of the
energy culumant (Fig. 3a) is close to the value expected of a continuous
transition, $\frac23$, indicating that it is only weakly first order.
This conclusion is corroborated by the estimated latent heat given in I.

\subsection{H=0.7: $6-7$ Transition}

The $6-7$ boundary is expected to be a line of continuous transitions
belonging to the Ising universality class.
MC histograms were made for $L=12-30$ at
several temperatures between $T=1.40$ and $T=1.45$ using $1-3 \times 10^6$
steps for averaging.  The transition temperature was estimated to be
$T_c=1.425(4)$.  Much difficulty was experienced
in locating extrema in the thermodynamic quantities for $L=12$ and these
data were not used in the finite-size scaling analysis.  Several runs using
the larger lattices exhibited widley varying results, indicating that critical
fluctuations are significant at this transition (possibly due to the
proximity of the paramagnetic state).  For example,
it was not possible to obtain
reliable results for the order parameter.  Fig. 4a shows a logarithmic
plot of the energy cumulant.  The extrapolated value $U^\ast=0.666665(4)$
suggests that the transition is
indeed continuous.  The slope $-2.9(1)$ is close to $-3$, as expected for
continuous transitions if the exponent ratio $\alpha / \nu$
is small,\cite{challa} as in the present case.  Finite-size scaling of the
other thermodynamic quantities shown in Fig. 4 was made with the assumption of
Ising critical exponents.\cite{ising}  As in the case of Fig. 3, these results
suggest that the true critical behavior is revealed only at larger lattice
sizes (especially the data for $V_{max}$).
Scaling with the assumption of volume dependence gave very
unsatisfactory results.

\subsection{H=2.7: $6-9$ Transition}

The anticipated difficulty in obtaining reliable estimates for the location of
extrema in the thermodynamic quantities as a function of temperature
at the $6-9$ transition was
realized.  This was due to the near zero slope of this boundary line
in the $H-T$ plane so that large fluctuations between the two phases
occurred.  (A more suitable, but a more memory- and
time-consuming, approach would have been
to obtain extrema as a function of magnetic field from the
histograms.\cite{ferr})  For this reason, finite-size scaling was performed
{\it at} the (crudely) estimated critical temperature.  Smaller lattice sizes,
$L=12-27$, were chosen in order to minimize the effect of an error in the
estimated $T_c$.  $2-7 \times 10^6$ MC steps were used to create histograms
at a number of temperatures between $T=1.20$ and $1.28$.  Scaling consistent
with the anticipated Ising universality, as shown in Fig. 5, was found by
assuming a critical temperature $T_c=1.23(1)$.  Note that the scale of
Fig. 5a for the specific heat is approximately the same as used in Fig. 4b
and reflects the very small slope of the fitted line.
Our MCH results for the order parameter were again found not to be
reliable for this transition.
The extrapolated value of the energy cumulant, $U^\ast=0.6666668(8)$,
clearly indicates that the transition is continuous.  As in the previous case,
the slope $-2.99(5)$ of Fig. 5b is consistent with a small value for
$\alpha / \nu$.

\subsection{H=4.0: $1-9$ Transition}

Although the $1-9$ transition is expected to belong to the XY universality
class within mean-field theory, the not unlikely possibility exists that
the cubic term in the free energy becomes relevant when critical fluctuations
are included.  Histogram data were taken at $T=1.42$ and $T=1.44$ on lattice
sizes $L=12-33$ with
$1-3 \times 10^6$ steps for averaging. Volume-dependent scaling was not
observed in the extrema of thermodynamic functions.  Further support for
the continuous nature of this transition comes from the extrapolated
value $U^\ast = 0.6666664(8)$, where the energy-cumulant scaling exponent
was found to be $-2.8(2)$.  Finite-size scaling consistent with XY
universality is shown in Fig. 6, including the order parameter evaluated
at the estimated critical temperature $T_c=1.423(4)$.

\section{Conclusions}

This work on the phase diagram of ferromagnetically coupled antiferromagnetic
XY triangular layers contains two results of importance.  The first is that
although molecular-field theory fails,
a more general phenomenological model is capable of capturing all the
essential features.  This mean-field model also allows specific
predictions to be made regarding the expected critical behavior.
Secondly, the histogram Monte Carlo method has been demonstrated to be
useful in verifying this anticipated criticality through finite-size
scaling of thermodynamic functions.  Relatively large fluctuations observed
in the MC data is attributed to the frustration inherent in the triangular
geometry.  These results serve to complement and extend our earlier
work,\cite{plum} as well as that of Lee {\it et al.}\cite{lee} who studied
the corresponding two-dimensional system.  Although these authors suggest
nonuniversal critical behavior on all transition lines, except the
$1-6$ boundary, in the 2D case, our analysis suggests that only the
$1-9$ transition should exhibit XY symmetry and the consequent KT-like
excitations associated with such nonuniversality.  It is of interest to
perform histogram MC simulations on the 2D system to examine these issues
further.  As mentioned in I, a good experimental candidate for the present
3D model is $La_2Co_{1.7}$.\cite{gig}

\acknowledgements
This work was supported by NSERC of Canada and FCAR du Qu\'ebec.

\begin{figure}
\caption{Phase diagram from Ref. 6.
Indicated are the paramagnetic phase $1$, linear phases $6$ and $9$, and
the elliptical (chiral) phase $7$.
Solid and broken lines are guides to the eye and
indicate first and second-order transitions, respectively.
Solid circles at H=0.7, 1.5,
2.7, and 4.0 denote points at which finite-size scaling analyses were
performed (where results at H=0.7 for the $1-6$ transition are in
Ref. 6).}
\label{fig1}
\end{figure}

\begin{figure}
\caption{Mean-field results for the phase diagram, where (a) is based on
a free energy derived from molecular-field theory,
(b) is from the phenomenological
Landau free energy (3) with all parameters as in (a) except $B_4=1$, (c)
is from the same model as (b) but with $B_4=0.1$.}
\label{fig2}
\end{figure}

\begin{figure}
\caption{Finite-size scaling of extrema for the $1-6$ transition at $H=1.5$
near $T_c \simeq 1.522$ with the assumption of volume dependence.
(a) shows the energy cumulant, where values for $L=12$ and $15$ have
been omitted to allow for an expanded scale.  All results for $L=12-33$
are shown in (b) and (c) for the specific heat, susceptibility, and
logarithmic derivative of the order parameter.}
\label{fig3}
\end{figure}

\begin{figure}
\caption{Finite-size scaling of extrema for the $6-7$ transition at $H=0.7$
near $T_c \simeq 1.425$ with the assumption of Ising univerality using
data from lattices $L=15-30$.}
\label{fig4}
\end{figure}

\begin{figure}
\caption{Finite-size scaling results for the $6-9$ transition at $H=2.7$
of thermodynamic functions evaluated {\it at} $T_c \simeq 1.23$
with the assumption of Ising univerality using
data from lattices $L=12-27$.}
\label{fig5}
\end{figure}

\begin{figure}
\caption{Finite-size scaling of extrema for the $1-9$ transition at $H=4.0$
near $T_c \simeq 1.423$ with the assumption of XY univerality using
data from lattices $L=12-33$.}
\label{fig6}
\end{figure}

\end{document}